\documentclass[12pt,twoside]{amsart}
\usepackage{amssymb}
\usepackage[english]{babel}
\usepackage[utf8]{inputenc}
\usepackage{algorithm}
\usepackage[noend]{algpseudocode}
\usepackage{natbib}
\usepackage{comment} 
\usepackage{tikz}
\usetikzlibrary{calc}
\usepackage{verbatim}
\usepackage{subcaption}
\usepackage[labelformat=parens,labelsep=quad,skip=3pt]{caption}
\usepackage{graphicx}
\usepackage{enumerate}

\usepackage{kantlipsum} 
\def\EE{\mathbb{E}}    

\def\proof{\noindent{\em Proof.}~}
\def\eproof{\mbox{\ }\hfill$\square$}
\newtheorem{theorem}{Theorem}
\newtheorem{lemma}{Lemma}
\newtheorem{proposition}{Proposition}
\newtheorem{corollary}{Corollary}
\newtheorem{assumption}{Assumption}
\newtheorem{definition}{Definition}

\newtheorem{example}[theorem]{\textsc{Example}}
\newtheorem*{example*}{Example}
\date{}

\title{Wishful Thinking is Risky Thinking}
\author{Jarrod Burgh\quad \quad Emerson Melo}

\date{\today}

\calclayout
\calclayout
\thanks{Department of Economics,
	Indiana University, Bloomington, IN 47408, USA. Email: {\tt \{jburgh,emelo\}@iu.edu}. We are very grateful to Duarte Goncalves, Marco Acosta, Jack Berger, Collin Raymond, Ben Bushong,  and participants at SAET 2023 and Georgia-Tech MWET 2023, for their  valuable comments and suggestions that have greatly improved the paper. This paper, previously circulated under the title ``\emph{Wishful Thinking is Risky Thinking: A Statistical-Distance Based Approach}," has been divided into two separate and complementary works: ``\emph{Wishful Thinking is Risky Thinking}'' and ``\emph{Censored Beliefs and Wishful Thinking.}''}
 
\begin{document}
	
	\maketitle
	
	\pagestyle{myheadings} \thispagestyle{plain} \markboth{ }{ }
	\begin{abstract}  
We develop a model of wishful thinking  that incorporates the costs and benefits of biased beliefs.  We establish the connection between distorted beliefs and risk, revealing how wishful thinking can be understood in terms of risk measures. Our model accommodates extreme beliefs, allowing wishful-thinking decision-makers to assign zero probability to undesirable states and positive probability to otherwise impossible states. 
\end{abstract}

	\vspace{3ex}

	\thispagestyle{empty}
	
	\newcommand{\spacing}[1]{\renewcommand{\baselinestretch}{#1}\large\normalsize}
	\textwidth      5.95in \textheight 600pt
	\spacing{1.1}
	\newpage
	\section{Introduction}\label{Intro}
Wishful thinking (WT) behavior is the tendency of agents to overestimate the probability of favorable outcomes while underestimating the likelihood of unfavorable outcomes (\cite{Aue_et_al_2012}). In economic contexts, WT-biased beliefs foster overconfidence and optimism in decision-making  (\cite{puri2007optimism}, \cite{Malmendier_Taylor_2015}, and \cite{Benabou_Tirole_2016}).
\smallskip 

There is substantial evidence of WT behavior in economic decision-making, with studies spanning various domains to illustrate this phenomenon. For instance, \cite{Oster_et_al2013} demonstrate how individuals at risk for Huntington's disease display optimism by deciding not to undergo genetic testing for early detection. \cite{Engelman_et_al_2024} provide evidence that people engage in WT to alleviate anxiety about adverse future outcomes. \cite{orhun2021motivated} analyze beliefs about the risks of returning to work during a pandemic. \cite{Malmendier_Taylor_2015_second}  offer insights into CEOs' overconfidence, while \cite{Kent_Hirshleifer_2015} discuss WT's role in explaining investors' optimistic behavior in financial markets. 
\smallskip

In this paper, we introduce a model of WT that comprehensively considers both the benefits and drawbacks associated with optimistic biased beliefs and the behavior associated with these beliefs. Drawing upon the frameworks established by \cite{BRACHA201267} and \cite{CaplinLeahy2019}, we propose a two-stage model where the decision maker (DM) confronts uncertainty regarding future events and makes decisions involving actions and belief structures. Through actively shaping their beliefs, the DM aims to maximize subjective utility for a given alternative while accounting for costs associated with deviating from prior beliefs.

To quantify this cost, we introduce a belief distortion function that is proportional to the $\phi$-divergence (\cite{Csiszar1967}) between subjective and prior beliefs. This cost function penalizes deviations from prior beliefs by measuring the statistical distance between subjective and prior beliefs. Our approach encompasses various examples, including the Kullback-Leibler, Burg, and modified $\chi^2$ distances. From a technical standpoint, our emphasis is on the category of $\phi$-divergence functions that are strictly convex and twice differentiable.
\smallskip

 We contribute to the literature on WT by making the following contributions. Firstly, we establish a direct link between WT behavior and the concept of convex risk measures, as introduced by \cite{Artzner_et_al_1999} and \cite{FRITTELLI20021473}. Using Lagrangian duality arguments, we demonstrate that the problem of selecting an optimal belief vector can be reformulated as a \emph{dual} minimization problem, which determines the level of risk associated with the optimal belief vector. Specifically, for the Kullback-Leibler distance, our analysis reveals that the WT problem corresponds to the DM analyzing the well-known entropic risk functional, as discussed in the works of \cite{Follmer_Schied_2002} and \cite{FollmerSchied_2016}. This behavioral perspective highlights that optimistic DMs tend to adopt beliefs that lead them to behave as risk seekers.
\smallskip

Second, we provide a complete characterization of optimal beliefs in the context of WT and offer new behavioral insights. We show that optimal beliefs exhibit a distinct pattern of ``twisting" prior probabilities toward states with high utilities. Importantly, this finding extends beyond the Kullback-Leibler case and applies to the broad class of $\phi$-divergence functions, going beyond previous research by \cite{CaplinLeahy2019} and \cite{mayraz2019priors}.
\smallskip

Third, we discuss how our model can capture situations where an optimistic DM assigns a subjective probability of zero to states with low utilities, reflecting a phenomenon we refer to as ``\emph{cognitive censoring}." We establish the necessary conditions on the cost function to capture this behavioral bias. Similarly, we identify necessary conditions that allow WT to lead the DM to assign positive subjective probabilities to states that the prior deems impossible or highly improbable, a behavior we term ``\emph{cognitive emergence}." Notably, we show that cognitive emergence is only observed for the state with the highest utility level. Our formalization of cognitive censoring and cognitive emergence aligns with concepts found in the psychology literature, such as ``wishing" (\cite{bury2016giving}) and ``false hope" (\cite{korner1970hope}), and sheds new light on these cognitive biases by emphasizing the significant role that subjective beliefs play in reshaping the perception of what is possible or not. To our knowledge, cognitive censoring and cognitive emergence represent novel contributions to the WT literature.
\smallskip

The rest of the paper is organized as follows: in \S \ref{model}, we introduce the model and characterize the optimal beliefs. Furthermore, in \S \ref{model} we discuss the connection between risk measures and WT models. \S \ref{extremesection} discusses how our  model can generate cognitive emergence and censorship.  \S \ref{relatedlit} discusses some of the related literature.  \S\ref{concusions} concludes.  Proofs and technical discussions are gathered in Appendices  \ref{AppendixA} and \ref{details_examples}, respectively.
\section{The Model}\label{model} 
In this section, we develop a WT model that incorporates the benefits and costs of distorted beliefs. As the introduction section mentions, we build upon  \cite{bracha2012affective} and  \cite{CaplinLeahy2019}'s framework.\footnote{ Similar to \cite{BrunnermeierParker_20005},  \cite{Benabou2013}, and \cite{CaplinLeahy2019}, we follow the  idea that the DM   maximizes her current subjective expected utility, which incorporates utility from current experience (assumed zero) and utility from the DM's anticipated future realization (anticipatory utility). This relies on the view that an agent's subjective utility depends on beliefs regarding future outcomes.}
\smallskip

Formally, we consider an environment where the DM is confronted with the task of selecting an action $a$ from a set $A=\{a_1,\ldots,a_n\}$ in the presence of uncertainty regarding a utility-relevant state $\omega \in \Omega$. We assume that the state space $\Omega$ is finite. The DM's utility function is defined as $u: A\times \Omega\longrightarrow \mathbb{R}$, which assigns a real-valued number to each pair $(a,\omega)$. Accordingly, we define the vectors  $U(a)=\left(u(a,\omega)\right)_{\omega\in \Omega}$ and $U=(U(a))_{a\in A}$ respectively. An \emph{exogenous} prior  belief is presented in the form of a probability distribution $q$ over the state space $\Omega$. The prior belief assigns a probability $q(\omega)\geq 0$ to each state $\omega\in \Omega$. The prior can be viewed as representing an objective likelihood of a state occurring, a data-driven probability based on similar experiences, or the beliefs of an expert.
The DM holds a subjective belief represented by the probability distribution $p \in \Delta(\Omega)$, where $p(\omega)$ denotes the subjective probability assigned to each state $\omega \in \Omega$. Intuitively, $p$ can be interpreted as a distortion of $q$.\footnote{ As we shall see, in general the subjective belief $p$ will depend in the optimal action $a\in A$.}

The subjective expected utility (SEU) of alternative $a \in A$ for the DM is given by:
\begin{equation}\label{Wishful_Thinking_EU}
   \mathbb{E}_{p}(u(a,\omega))=\sum_{\omega \in \Omega} p(\omega) u(a,\omega)
\end{equation}

The expected payoff (\ref{Wishful_Thinking_EU}) makes explicit that the DM uses her subjective beliefs $p$ to evaluate  utility-maximizing actions.  
\smallskip

To account for the impact of deviating from the beliefs $q$, we introduce a cost of belief distortion in evaluating utility-maximizing actions. This cost reflects the DM's preference for accurate beliefs. We assume belief distortion costs increase as the deviation from $q$ increases. Formally, we model the cost of belief distortion as the $\phi$-divergence (\cite{Csiszar1967}) between the subjective belief $p$ and $q$, denoted as $C_{\phi}(p\| q)$. This cost function captures the \emph{statistical distance} between subjective and prior beliefs. The details of $C_{\phi}(p\| q)$ will be discussed in the subsequent section.

\smallskip


Accordingly, and given a cost function $C_{\phi}(p\|q)$, the WT agent chooses an optimal pair $(a^\star,p^\star)$ that maximizes:
\begin{equation}\label{wishful_thinking_problem}
  \max_{a\in A}\max_{p\in\Delta(\Omega)}  \left\{\EE_p(u(a,\omega)) - \delta C_{\phi}(p\|q)\right\}
\end{equation}
where $\delta>0$ represents the marginal cost associated with deviations from $q$.

\smallskip 

The problem (\ref{wishful_thinking_problem}) presents a framework in which the agent simultaneously chooses both an action and a belief structure. Importantly, this framework considers the subjective beliefs to be contingent on the chosen action. In other words, the DM assigns a specific belief structure to each possible action, allowing for the possibility of holding seemingly contradictory beliefs. This notion of beliefs being action-contingent is reminiscent of the concept of cognitive dissonance, as discussed in \cite{akerlof1982economic} and relates to situations where agents may hold contradictory beliefs. For an illustrative example clarifying this concept of action contingency, please refer to Section \ref{actioncont}.
\smallskip

Our approach to WT behavior is closely connected to motivated reasoning, as discussed in studies such as \cite{Kunda1990TheCF} and \cite{Benabou_Tirole_2016}. According to this theory, when choosing their optimal beliefs, a motivated DM is driven by the subjective expected utility $\EE_p(u(a,\omega))$, representing the anticipatory expected utility associated with different actions and outcomes. However, in addition to the anticipatory utility, a motivated DM also considers the importance of accuracy. In our framework, this is captured by the term $\delta C_\phi(p\|q)$ in expression (\ref{wishful_thinking_problem}), where $\delta$ represents the weight placed on the cost of belief distortion. Thus, our model incorporates the motivational aspect of maximizing SEU and considering accuracy in belief formation.

\subsection{Belief distortion and $\phi$-divergences}\label{beliefcost} Intuitively, the term $C_\phi(p\|q)$ captures the distance or divergence between  $p$ and $q$. We formalize this interpretation by employing the concept of statistical divergence, which is a measure of dissimilarity between probability distributions (\cite{Csiszar1967}; \cite{LieseVajda1987}; \cite{Pardo2005}). To do so, we focus on a specific class of  $\phi$-divergence functions. 
\begin{definition}
Consider the class of $\phi$-divergence functions $\Phi$. A function $\phi \in \Phi$ must satisfy: 
    \begin{enumerate}
        \item $\phi: \mathbb{R} \rightarrow(-\infty,+\infty]$ is a proper closed convex function
        \item $\phi$ is non-negative and  attains its minimum at  $1$, and furthermore $\phi(1) = 0$.
        \item Define undefined arguments as: $0\phi(\frac{c}{0}) = c lim_{t \rightarrow \infty} \frac{\phi(t)}{t}$ $\forall c> 0$ and $0\phi(\frac{0}{0}) = 0$.
    \end{enumerate}
\end{definition}

 We can now utilize our definition of the class $\Phi$ to formalize our cost function:

\begin{definition}\label{Phi_Definition}
	Let $\phi \in \Phi$. The $\phi$-divergence of the probability vector $p$ with respect to the prior belief $q$ is
	\begin{equation} \label{phi_divergence_definition}
		C_\phi(p\|q)=\sum_{\omega\in \Omega}q(\omega)\phi\left(\frac{p(\omega)}{q(\omega)}\right).
	\end{equation}
	where $p, q \in \Delta(\Omega)$.	
\end{definition}
In equation (\ref{phi_divergence_definition}), the cost function explicitly depends on the choice of $\phi\in \Phi$. Additionally, we observe that $C_{\phi}(p\|q)$ can be expressed as the expected value under the prior $q$ of the function $\phi$ applied to the ratio $p(\omega)/q(\omega)$ for all $\omega\in  \Omega$.

While $C_\phi(p\|q)$ is a statistical distance, it is worth mentioning that it does not necessarily satisfy the triangle inequality. Moreover, for $p$ and $q$ in the interior of the probability simplex, it is generally true that $C_{\phi}(p
\|q)\neq C_{\phi}(q\|p)$. In other words, the function $C_\phi(\cdot\|\cdot)$ is generally asymmetric.
\smallskip

A key element in our framework will be the convex conjugate of the function $\phi$, denoted as $\phi^\ast$. Formally, the conjugate function $\phi^\ast$ is defined as:
\begin{equation}\label{Phi_convex_conjugate}
\phi(s)=\sup _{t \in \mathbb{R}}{s t-\phi(t)}=\sup _{t \in \operatorname{dom} \phi}{s t-\phi(t)}=\sup _{t \in \operatorname{int} \operatorname{dom} \phi}{s t-\phi(t)},
\end{equation}
where the last equality follows from \cite[Cor. 12.2.2]{Rockafellar1970}. The conjugate function $\phi^\ast$ is a closed proper convex function, with $\operatorname{int} \operatorname{dom} \phi^\ast=(a, b)$, where $$
 a=\lim _{t \rightarrow-\infty} t^{-1} \phi^\ast(t) \in[-\infty,+\infty) ; b=\lim _{t \rightarrow+\infty} t^{-1} \phi^\ast(t) \in(-\infty,+\infty] .
 $$
 
Additionally, it is essential to note that for the convex and closed function $\phi$, its bi-conjugate is given by $\phi^{* *}=\phi$, as shown in \cite{Rockafellar1970}. 

A key observation is that since $1$ is the minimizer of $\phi$ and lies in the interior of its domain, we have $\phi'(1) = 0$. Moreover, utilizing the property of convex and closed functions, known as the Fenchel equality, we have the equivalence $y =\phi'(x)$ if and only if $x = \phi^{*'}(y)$. By applying this observation to $x=1$ and $y=0$, we obtain $\phi^{\ast\prime}(0)=1$.
 \smallskip
 
Throughout the paper, we make the following assumption.
 \begin{assumption}\label{phi_diffferentiable_Assumption}$\phi^*(s)$ is strictly convex and differentiable with $\phi^{*\prime}(s)\geq0$ for all $s$. 
\end{assumption}
\scriptsize{
\begin{center}
\begin{table}
\begin{tabular}{||c c c c||} 
 \hline
  Divergence & $\phi(t)$ & $C_\phi(p\|q)$ & $\phi^*(s)$ \\ [0.5ex] 
 \hline\hline
 Kullback-Leibler & $\phi(t):=t \log t-t+1, \quad t>0$ & $\sum_{\omega\in \Omega} p(\omega) \log \frac{p(\omega)}{q(\omega)}$ & $e^s-1$ \\ 
 \hline
 Hellinger & $(1-\sqrt{t})^2, \quad t>0$ & $\sum_{\omega\in \Omega}\left(\sqrt{p(\omega)}-\sqrt{q(\omega}\right)^2$ & ${s\over 1-s} \quad s<1$\\
 \hline
 Modified $\chi^2$ distance & $\phi(t):=(t-1)^2, \quad t>0$ & $\sum_{\omega\in \Omega} \frac{\left(p(\omega)-q(\omega)\right)^2}{q(\omega)}$ & $\phi^\ast(s)=\begin{cases}
			-1, & \text{if $s<-2$ }\\
            s+{s^2\over 4}, & s\geq -2.
		 \end{cases}$\\
\hline
Burg Entropy & $-logt + t - 1, \quad t> 0$ & $\sum_{\omega\in \Omega} q(\omega) \log \frac{p(\omega)}{q(\omega)}$ & $\phi^*(s) = -log(1-s), \quad s< 1$\\
\hline
\end{tabular}
\caption{Example of $\phi$-divergences and their conjugates.}
\label{table:1}
\end{table}
\end{center}}
\normalsize
Table \ref{table:1} provides some popular examples of $\phi$-divergences and their conjugates. As it is easy to see, the Kullback-Leibler distance is a particular case of a wider class of tractable cost functions. Furthermore, the conjugate $\phi^\ast$ has a very tractable form for these four cost functions.  As we shall see, this later property will be useful in characterizing the optimal beliefs.  In doing so, we make use of the following technical lemma.
 
\begin{lemma}\label{dualprob} Let Assumption \ref{phi_diffferentiable_Assumption} hold.  Let $V_\phi(U(a))\triangleq\max_{p\in\Delta(\Omega)}\left\{\mathbb{E}_{p}(u(a,\omega))-\delta C_\phi(p\|q) \right\}$ for all $a\in A$. Then, the following holds:
\begin{equation}\label{equivalence_belief_choice}
V_\phi(U(a))=\min_{\lambda_a\in[\underline{u}_a,\bar{u}_a]} \left\{ \lambda_a+\delta \EE_{q}(\phi^\ast((u(a,\omega)-\lambda_a)/\delta))\right\},
\end{equation}where $\underline{u}_a=\min_{\omega\in \Omega}{u(a,\omega)}$  and  $ \bar{u}_a=\max_{\omega\in \Omega}u(a,\omega)$. Furthermore, the minimization problem (\ref{equivalence_belief_choice}) has a unique optimal solution $\lambda_a^\star$.
\end{lemma}
\proof All proofs are gathered in Appendix \ref{AppendixA}.\eproof

\smallskip
 The previous lemma establishes that finding the optimal beliefs is equivalent to solving a one-dimensional minimization problem.

In what follows, let $\lambda_a^\star$ be the unique solution to the problem (\ref{equivalence_belief_choice}). By using Lemma \ref{dualprob}, we are able to characterize the DM's optimal belief choice vector.

\begin{proposition}\label{Belief_choice_characterization} Let Assumption \ref{phi_diffferentiable_Assumption} hold and define $w_a(\omega)\triangleq\phi^{\ast\prime}((u(a,\omega)-\lambda^\star_a)/\delta)$ for all $a\in A,\omega\in \Omega$. Then  for each $a\in A$ the optimal  belief choice  $p^\star(a)$ satisfies 
\begin{equation}\label{Optimal_beliefs_gradient}    
\nabla V_\phi(U(a))=p^\star(a),
\end{equation}
  where
 \begin{equation}\label{Optimal_beliefs}
 {\partial V_\phi(U(a)) \over\partial u(a,\omega) }=w_a(\omega)q(\omega)=p_a^\star(\omega), \quad \forall \omega\in \Omega.
 \end{equation}
\end{proposition}

Some remarks are in order. First, Proposition \ref{Belief_choice_characterization} characterizes the optimal belief vector $p^\star(a)=(p_a^\star(\omega))_{\omega\in \Omega}$ as the product of the weight vector $w_a=(w_a(\omega))_{\omega\in \Omega}$ and the prior belief $q$. Additionally, from the definition of $w_a(\omega)$, the distorted belief vector $p^\star(a)$ depends on the particular choice of $\phi$. Each term $w_a(\omega)$ captures how the DM ``twists" the truth (\cite{kovach2020twisting}) associated withe the probability of observing  the state $\omega$. For instance, if $w_a(\omega)=1$ for some $\omega\in \Omega$, then $p^\star_a(\omega)=q(\omega)$. Similarly, if $w_a(\omega)>1$ for some $\omega\in \Omega$, then $p^\star_a(\omega)>q(\omega)$. In the latter case, we say that the DM exhibits overprecision (\cite{Moore_et_al_2015}).\footnote{It is noteworthy that we can relax Assumption \ref{phi_diffferentiable_Assumption} by considering the subgradient $\partial \phi^\ast(s)$ instead of the gradient $\phi^{\ast\prime}$(s). Due to the convexity of $\phi^\ast$, the subgradient $\partial \phi^\ast(s)$ always exists. Consequently, the weight $w_a(\omega)$ corresponds to a selection of $\partial \phi^\ast(s)$. Moreover, given that the subgradient of a convex function is a maximal monotone operator, the monotonicity result in Corollary \ref{higherstate} will also remain valid.}
\smallskip

Second, the result in Proposition \ref{Belief_choice_characterization} is related to \cite{mayraz2019priors}. In his paper, the weights $w_a(\omega)$ are interpreted as ``desires" that capture the DM's overoptimism. Specifically, \cite{mayraz2019priors} models the weights $w_a(\omega)$ as proportional to $e^{u(a,\omega)/\delta}$. Therefore, our characterization (\ref{Optimal_beliefs}) generalizes the incorporation of desires in WT models. In addition, the characterization in Proposition \ref{Belief_choice_characterization} generalizes the result in \cite{CaplinLeahy2019} non-trivially. They characterize the optimal beliefs $p^\star(a)$ under the assumption that $C_{\phi}(p\|q)$ is given by the Kullback-Leibler distance. Their characterization is a particular case of the expression (\ref{Optimal_beliefs}). 
\smallskip

Third, the expression (\ref{Optimal_beliefs}) helps us to understand how our model captures WT behavior in a  general way. To see this, let us consider the ratio of the optimal beliefs for two states and assume that for $\omega,\omega^\prime\in \Omega$ the associate utilities satisfy $u(a,\omega)>u(a,\omega^\prime)$. Then  for  $q(\omega)>q(\omega^\prime)$, the likelihood ratio

\begin{equation}\label{log_odd_ratio}
    {p^\star_a(\omega)\over p^\star_a(\omega^\prime)}={w_a(\omega)q(\omega)\over  w_a(\omega^\prime)q(\omega^\prime)}
\end{equation}
implies that $p_a^\star(\omega)> p^\star_a(\omega^\prime).$  
\smallskip

The likelihood ratio provides a more precise understanding of the behavioral implications of  WT subjective beliefs. Formally, the ratio in (\ref{log_odd_ratio}) encapsulates the notion that when comparing two states, $\omega$ and $\omega^\prime$, the relative probability assigned to $p_a^\star(\omega)$ is higher than $p_a^\star(\omega^\prime)$. Consequently, the DM assigns a higher probability to more desirable outcomes in relative terms. In other words, the DM's optimal belief vector $p^\star(a)$  biases the prior  $q$ towards states with higher utilities. This optimistic biased behavior holds for all $\phi\in \Phi$. The following corollary formalizes the preceding discussion.
\begin{corollary}\label{higherstate} Let Assumption \ref{phi_diffferentiable_Assumption}  hold. Then $p_a^\star(\omega)$ is increasing in both $u(a,\omega)$ and $q(\omega)$. Therefore, given states $\omega$ and $\omega^\prime$ with $q(\omega) > q(\omega^\prime)$  and $u(a,\omega)>u(a,\omega^\prime)$, we have $p_a^\star(\omega)>p_a^\star(\omega^\prime)$.
\end{corollary}


\begin{example}\label{WTKLexample}
To comprehend how Proposition \ref{Belief_choice_characterization} operates, let us revisit the scenario where $C_\phi(p| q)$ represents the Kullback-Leibler distance. It is evident that $\lambda^\star_a=V_{\phi}(U(a))=\delta\log\mathbb{E}_q(e^{u(a,\omega)/\delta})$. Therefore, following Proposition \ref{Belief_choice_characterization}, expression (\ref{Optimal_beliefs}) indicates that subjective beliefs are determined by:

    $$p_a^\star(\omega)=w_a(\omega)q(\omega)$$
    where $$w_a(\omega)={\frac{e^{u(a,\omega)/\delta}}{\sum_{\omega^\prime \in \Omega} q(\omega^\prime)e^{u(a,\omega^\prime)/\delta}}}$$
Consequently, the DM selects the optimal action $a^\star\in A$ such that:
$$a^\star = \arg\max_{a\in A}\delta\log(\mathbb{E}_q(e^{u(a,\omega)/\delta})).$$\eproof
\end{example}

\begin{example}\label{Renyi2Example}
Let the DM solve the problem (\ref{wishful_thinking_problem}), where the modified $\chi^2$ distance gives the deviation cost from $q$. Assume
$\delta>( \mathbb{E}_q(u(a,\omega)) -\underline{u}_a )/2$ for all $a\in A$ and $\omega\in \Omega$. It is straightforward to show that $$\lambda_a^\star = \mathbb{E}_q(u(a,\omega)).$$

For a given action $a$, let us define the following difference:
$$g(a,\omega) \triangleq u(a,\omega) - \mathbb{E}_q(u(a,\omega)).$$

Accordingly, we can then write $$V_{\phi}(U(a))= \mathbb{E}_q(u(a,\omega)) + {1\over 4\delta}\sum_{\omega \in \Omega}q(\omega)g(a,\omega)^2.$$

Defining $Var(U(a))\triangleq \EE_q(g(a,\omega)^2) $, we rewrite $V_{\phi}(U(a))$ as
$$V_{\phi}(U(a))=\EE_q(u(a,\omega))+{1\over 4\delta}Var(U(a)).$$

Thus, the optimized valued associated with  $p^\star(a)$ takes the form of a mean-variance model. To recover the subjective beliefs, we use Proposition \ref{Belief_choice_characterization} and compute $ {\partial V_\phi(U(a)) \over\partial u(a,\omega) }$:
\begin{center}
$ {\partial V_\phi(U(a)) \over\partial u(a,\omega) } = q(\omega) + \frac{1}{2\delta}q(\omega)g(a,\omega) - \frac{1}{2\delta}q(\omega)\sum_{\omega^\prime \in \Omega} q(\omega^\prime) g(a,\omega^\prime)$
\end{center}

Using the fact that 
$\sum_{\omega \in \Omega} q(\omega) g(a,\omega) = 0$, the subjective beliefs are:

$$p_a^\star(\omega) = q(\omega)\left(1 + \frac{g(a,\omega)}{2\delta}\right)\quad \forall \omega\in \Omega.$$

The previous expression shows that  $p_a^\star(\omega)$ is increasing in $g(a,\omega)$.  Finally,  the DM chooses the optimal action $a^\star\in A$ such that:

$$a^\star = \arg \max_{a \in A}\left\{ \EE_q(u(a,\omega))+{1\over 4\delta}Var(U(a))\right\}.$$\eproof

\end{example}

In Example \ref{Renyi2Example}, the DM prefers actions associated with both a high mean and a high variance. This preference is reflected in the optimal belief vector $p^\star(a)$.

By examining the likelihood ratio, we can further explore the subjective probabilities. In the case of the Kullback-Leibler distance, the likelihood ratio is given by:

\begin{equation}
\frac{p_{a^\star}^\star(\omega)}{p_{a^\star}^\star(\omega^\prime)} = \frac{q(\omega)}{q(\omega^\prime)}e^{u(a^\star,\omega)-u(a^\star,\omega^\prime)}.
\end{equation}

The likelihood ratio in the Kullback-Leibler case has been widely discussed in the WT literature, including works by \cite{CaplinLeahy2019}, \cite{mayraz2019priors}, and \cite{kovach2020twisting}. It captures the relative weighting of states $\omega$ and $\omega^\prime$ based on the prior probabilities $q$, as well as the differences in utilities $u(a^\star,\omega)-u(a^\star,\omega^\prime)$.

In the case of the modified $\chi^2$ distance, the likelihood ratio is given by:

\begin{equation}
\frac{p^\star_{a^\star}(\omega)}{p_{a^\star}^\star(\omega^\prime} = \frac{q(\omega)(g(a^\star,\omega) + 2\delta)}{q(\omega^\prime)(g(a^\star,\omega^\prime) + 2\delta)}.
\end{equation}

In this case, the likelihood ratio depends on the extent to which the utilities associated with states $\omega$ and $\omega^\prime$ are better or worse than the average utility.

It is important to note that closed-form solutions for the likelihood ratios are possible in the cases of KL and modified $\chi^2$ distances. However, in general, obtaining closed-form solutions is not always feasible. Analyzing the likelihood ratio provides valuable insights into how the prior beliefs and utility differences between states influence the DM's subjective probabilities.


\subsection{WT behavior and risk }   In the context of portfolio allocation, \cite{Follmer_Schied_2002} and \cite{FRITTELLI20021473} have introduced the notion of \emph{convex risk measures} in an attempt  to quantify the riskiness of different financial portfolio decisions.\footnote{In a fundamental paper, \cite{Artzner_et_al_1999} introduced  the notion of coherent risk measures. Their approach is axiomatic and relies heavily on the properties of subadditivity and homogeneity. \cite{Follmer_Schied_2002} and \cite{FRITTELLI20021473} replaced these two conditions by focusing in convexity properties. It is worth pointing out that the notion of coherent and convex risk measures applies far beyond the case of portfolio allocation problems.} In our WT model, the notion of convex risk measures emerges naturally to quantify the risk associated with choosing a pair $(a,p(a))$. Formally, $V_\phi$ possesses all the  defining properties of a convex risk measure. The following result formalizes this connection.
\begin{proposition}\label{Vproperties} Let $\phi \in \Phi$. Then for all $a\in A$, the following properties hold:
\begin{itemize}
\item[(i)] $V_\phi(U(a)+c)=V_\phi(U(a))+c, \forall c \in \mathbb{R}$.
\item[(ii)] $V_\phi(c)=c$, for any constant $c \in \mathbb{R}$ (considered as a degenerate random variable).
\item[(iii)]If $u(a,\omega) \leq \tilde{u}(a,\omega), \forall \omega \in \Omega$, then, $V_\phi(U(a)) \leq V_\phi(\tilde{U}(a))$. 
\item[(iv)] For any two random variables $U_1(a), U_2(a)$ with finite moments and any $\alpha \in(0,1)$, one has
$$
V_\phi\left(\alpha U_1(a)+(1-\alpha) U_2(a)\right) \leq \alpha V_\phi\left(U_1(a)\right)+(1-\alpha) V_\phi\left(U_2(a)\right).
$$
\end{itemize}

\end{proposition}

The previous result is a simple adaptation of \cite[Thm. 2.1 ]{BenTal_Teboulle_2007}. Part (i) is known as \emph{translation invariance} and establishes that  adding a constant $c$  to $U(a)$ is equivalent to adding the same constant to $V_\phi(U(a))$. Part (ii) is known as \emph{consistency} and states that when $u(a,\omega)=c$ for all $\omega$, then the value of $V_\phi(U(a))$ is  constant and equal to $c$. Part (iii) is just a \emph{monotonicity} condition, in the sense that  $V_\phi(U(a))$ is monotone increasing on $U(a)$ (in a stochastic sense). Finally, condition (iv)  establishes that $V_\phi(U(a))$ is a \emph{convex} function. 
\smallskip

As we said before, the properties in Proposition \ref{Vproperties} establish that  $V_\phi(U(a))$ is a convex risk measure. The following corollary formalizes this fact.

\begin{corollary}\label{V_risk_measure} For all $a\in A$, $V_\phi(U(a))$ is a convex risk measure.
\end{corollary}

Some remarks are in order. First, to see why the interpretation of $V_\phi$  as a risk measure is useful, we revisit the Kullback-Leibler case. In this case,  we know that $V_\phi(U(a))=\delta\log(\mathbb{E}_q(e^{u(a,\omega)/\delta}))$, which  is known as the entropic risk measure (\cite{FollmerSchied_2016}). Similarly, in the case of the Example \ref{Renyi2Example}, we know that for each alternative $a\in A$, we get  $V_\phi(U(a))=\EE_q(u(a,\omega))+{1\over 4\delta}Var(U(a))$. Noting that $\delta>0$, the DM will choose the alternative with the highest expected utility and variance combination. Corollary \ref{V_risk_measure} establishes that this pattern generalizes beyond the entropic and the mean-variance cases. However, it's important to highlight that in our WT model, the implementation of the notion of convex risk measure differs from its traditional application in portfolio allocation problems. The key distinction lies in the fact that in our model, each alternative $a\in A$ is associated with an uncertain prospect $U(a)$, and the associated risk is measured by $V_\phi(U(a))$. This framework allows us to analyze risk and decision-making under uncertainty in a more general context beyond traditional portfolio allocation settings.

\smallskip

Second, it is worth noting that Corollary \ref{V_risk_measure} implies that a WT agent will choose the riskiest alternative from the set $A$. This can be observed by combining problem (\ref{wishful_thinking_problem}) with Lemma \ref{dualprob}, which shows that the WT problem is equivalent to $\max_{a\in A}V_{\phi}(U(a))$. Thus, the WT agent aims to maximize the risk measure $V_{\phi}$ applied to the uncertain prospects $U(a)$ associated with each alternative $a\in A$. This behavior can be seen in Examples \ref{WTKLexample} and \ref{Renyi2Example}, where the WT agent selects the alternative with the highest risk according to the specified risk measure.
\smallskip

To our knowledge, the connection between risk measures and  WT behavior has not been  explored in the existing literature.

\subsection{ WT and EU behavior}
This section establishes the behavioral equivalence between WT decision-making and EU behavior. Recalling that $U=(U(a))_{a\in A}$ and $U(a)=(u(a,\omega))_{\omega\in \Omega}$, we define 
$$A_{EU}(U) \triangleq \arg \max_{a \in A} \mathbb{E}_q (u(a, \omega))$$
as the set of optimal actions associated with EU maximization.
\smallskip

Now, let  $\hat{u}(a,\omega)=\lambda_a^\star+\delta\mathbb{E}_{q}(\phi^\ast((u(a,\omega)-\lambda_a^\star)/\delta))$ and $\hat{U}(a)=(\hat{u}(a,\omega))_{\omega\in \Omega}$, and $\hat{U}=(\hat{U}(a))_{a\in A}$, for all $a\in A,\omega \in \Omega$. In addition, let
$$A_{WT}(U) \triangleq \arg\max_{a \in A}\max_{p\in\Delta(\Omega)}  \left\{\EE_p(u(a,\omega)) - \delta C_{\phi}(p\|q)\right\}.$$
 define the set of optimal actions for the WT problem. 
The following result establishes the equivalence between the EU and WT models. 
\begin{proposition}\label{euequiv}
A WT agent with utility function $u(a,\omega)$ is behaviorally equivalent to an EU maximizer agent with the transformed utility function $\hat{u}(a,\omega)$. In particular, 
$$A_{WT}(U) = A_{EU}(\hat{U})$$
\end{proposition}
Two remarks are in order. First, Proposition \ref{euequiv} establishes a general behavioral equivalence between WT models using $\phi$-divergences and EU maximization. It shows that WT behavior can always be interpreted as the outcome of EU maximization under a distorted utility function. An important implication of this equivalence is that we cannot distinguish between these models based on choice data alone. However, combining choice and belief data may determine which framework is more appropriate. This offers a potential avenue for practical model selection.

Second, the behavioral equivalence presented in Proposition \ref{euequiv} extends the findings of \cite{robson2022decision} non-trivially. The focus of \cite{robson2022decision} is on the specific case of Kullback-Leibler divergence, where the distorted utility function is given by $\tilde{u}(a,\omega) = \lambda_a^\star + \delta \exp\left(\frac{u(a,\omega) - \lambda_a^\star}{\delta}\right)$. By leveraging the relationship $\lambda_a^\star=\delta\log\mathbb{E}_{q}(e^{u(a,\omega)/\delta})$, it can be shown that $\mathbb{E}_q(\tilde{u}(a,\omega))=\delta\log\mathbb{E}_q(e^{u(a,\omega)/\delta})$, leading to $A_{eu}(\tilde{U})=\arg\max_{a\in A}\delta\log \mathbb{E}_q(e^{u(a,\omega)/\delta})$. The equivalence established in Proposition \ref{euequiv} goes beyond the specific case of Kullback-Leibler distance and holds for a broader class of $\phi$-divergences.

\subsection{WT as an intrapersonal game} We close this section discussing a final and  important implication of Lemma \ref{wishful_thinking_problem} and Proposition \ref{Belief_choice_characterization}. Together with these results, we  can represent optimal WT behavior as  the solution to a saddle point problem. The following result formalizes this observation. 

\begin{corollary}\label{saddleWT} For each $a\in A$, let $\Psi(a,\lambda_a)\triangleq \lambda_a+\delta\EE_q(\phi^*((u(a,\omega)-\lambda_a)/\delta)$. Then the pair $(a^\star,p^\star(a))$ solves (\ref{wishful_thinking_problem}) iff the pair  $(a^\star,\lambda_a^\star)$ solves the saddle point  problem:
$$\max_{a \in A}\min_{\lambda_a\in\Lambda(a)}\Psi(a,\lambda_a)$$
where $\Lambda_a\triangleq\{\lambda_a:min_{\omega\in \Omega}{u(a,\omega)} \leq \lambda_a\leq\max_{\omega\in \Omega}u(a,\omega)\}.$
\end{corollary}
The previous corollary is useful to understand WT decision-making in terms of a two-player intrapersonal game in the spirit of \cite{bracha2012affective}.  In their language, the player choosing the optimal action $a^\star\in A$ corresponds to the rational process while choosing the optimal belief $p^\star(a^\star)$ corresponds to the emotional process. In particular, the rational  agent solves $\max_{a\in A} \EE_{p}(u(a,\omega))$ while the emotional agent solves $\max_{p\in\Delta(\Omega)} \EE_p(u(a,\omega)) - \delta C_{\phi}(p\|q)$.
\smallskip

Intuitively, the  rational agent  is an EU maximizer, while the emotional agent is a subjective expected utility  maximizer with a taste for higher payoff states. \cite{bracha2012affective} show that the  objective function in our WT problem (\ref{wishful_thinking_problem}) is a potential function for this intrapersonal game. Thus, the solution of this intrapersonal game is identical to the solution to the problem (\ref{wishful_thinking_problem}). Corollary \ref{saddleWT} adds a useful interpretation, emphasizing that the tension between the rational and emotional agents  can be represented as a max-min problem incorporating an appropriate notion of risk. 

\section{Extreme  Beliefs}\label{extremesection}
In this section, we delve into how our WT model can capture ``extreme" beliefs by relaxing the assumption of absolute continuity, which states that a state has a zero prior probability if and only if it has a zero subjective probability. This assumption is commonly made in the literature on motivated reasoning, with a few exceptions, such as \cite{bury2016giving} and \cite{korner1970hope}.
\smallskip

We first examine optimal beliefs in the case of cognitive censorship, where the DM  subjectively ignores states with low payoffs. To exhibit this behavior, the DM selects a utility cutoff such that any state generating a utility below this cutoff is subjectively disregarded. This cognitive censorship reflects the DM's tendency to ignore or downplay unfavorable outcomes, focusing only on states with sufficiently high utilities. By doing so, the DM shapes their beliefs to align with their desired outcomes, exhibiting WT.
\smallskip

Next, we investigate optimal beliefs in the context of cognitive emergence, where the DM subjectively believes that the prior assigned probability zero (or an impossible state) can still be realized or observed with some positive subjective probability. However, this bias is only exhibited in states with the highest payoffs. In other words, the DM assigns positive subjective probabilities to these extreme states, despite their prior unlikelihood. This behavior reflects the DM's inclination to perceive even highly improbable outcomes as possible or likely when those outcomes align with their desired goals or aspirations.
\smallskip

By exploring these variations of belief formation, our WT model captures the cognitive processes of censorship and emergence. These behaviors highlight how the DM's subjective beliefs can deviate from prior probabilities, leading to extreme beliefs influenced by optimism.

\subsection{ Cognitive censorship} Optimal beliefs display cognitive censorship when, for some state $\omega\in \Omega$ with $q(\omega)>0$, the DM  chooses $p_{a^\star}^\star(\omega) = 0$.\footnote{In this section, for ease of notation, we set $\delta = 1$. Results can be easily extended to an arbitrary $\delta>0$.} 
\smallskip

We aim to characterize the class of $\phi$-divergence functions that implies  WT  behavior consistent with cognitive censorship.  The following proposition provides necessary conditions on $\phi$.\footnote{These conditions draw from similar results in a data-driven problem in \cite{bayraksan2015data}.} 
\begin{proposition}\label{zerochoice} Let $a^\star$ and $p^\star(a^\star)$ be an optimal solution to the WT problem (\ref{wishful_thinking_problem}). Then  $p^\star(a^\star)$  can generate cognitive censorship  only if
\begin{itemize}
\item[(i)] $\lim_{t \rightarrow 0+}\phi(t) < \infty$,
\item[(ii)] $\lim_{t \rightarrow 0+}\phi'(t) > -\infty$.
\end{itemize}

Furthermore, if these conditions hold, then there exists a cutoff $\tilde{u}(a^\star)$ such that $p_{a^\star}^\star(\omega) = 0$ if and only if $u(a^\star,\omega) \leq \tilde{u}(a^\star)$.
\end{proposition}

These conditions ensure that the $\phi$-divergence function exhibits the necessary behavior to accommodate corner solutions in the WT model. The boundedness of the cost function when $p_{a^\star}^\star(\omega)=0$ is crucial for such solutions, allowing the DM to selectively ignore certain states and assign them zero probabilities based on their utilities.

In summary, Proposition \ref{zerochoice} provides conditions on the $\phi$-divergence function that allows for  optimal beliefs with corner solutions, where some states are assigned zero probabilities. These conditions capture the DM's cognitive censorship behavior, where certain states are subjectively disregarded based on their utilities.
\smallskip

The second assertion of Proposition \ref{zerochoice} describes how the cognitive censorship behavior takes place in the WT model. The DM determines a cutoff value, denoted as $\tilde{u}(a^\star)$, representing the threshold utility below which states are subjectively censored or ignored. Intuitively, the DM starts by censoring the state with the lowest utility, then proceeds to censor the state with the second-lowest utility, and so on.

By setting the cutoff value $\tilde{u}(a^\star)$, the DM effectively ignores states that provide utilities below this threshold, treating them as having zero probability. This cognitive censorship behavior allows the DM to focus on states perceived as more desirable or relevant based on their utilities while disregarding states considered less favorable.

In summary, the second assertion of Proposition \ref{zerochoice} clarifies that in the context of cognitive censorship, the DM establishes a cutoff value $\tilde{u}(a^\star)$ and selectively censors states with utilities below this threshold. This behavior allows the DM to prioritize and focus on states deemed more favorable or significant while ignoring less desirable states.
\smallskip

To see how the result works, we analyze an environment  where $\Omega=\{ \omega_H,\omega_L\}$ and $A=\{1,\ldots,n\}$. Let $(a^\star,p^\star(a^\star))$ be an optimal solution where  the state contingent utilities  are $u(a^\star,\omega_H) = 4$ and $u(a^\star,\omega_L) = 0$. 

\begin{figure}
\caption{Subjective Probabilities vs. Prior Probabilities}
  \begin{subfigure}{6cm}
    \centering\includegraphics[width=5cm]{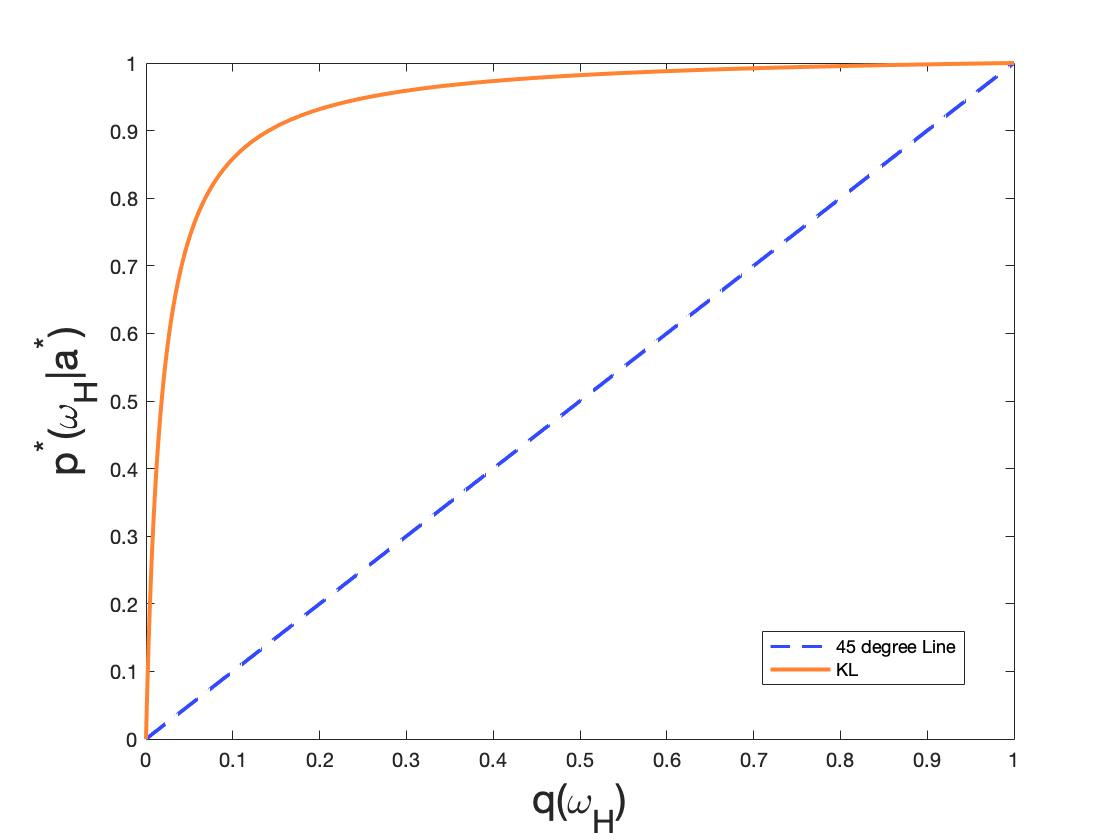}
    \caption{Kullback-Leibler}
    \label{KLgraph}
  \end{subfigure}
  \begin{subfigure}{6cm}
    \centering\includegraphics[width=5cm]{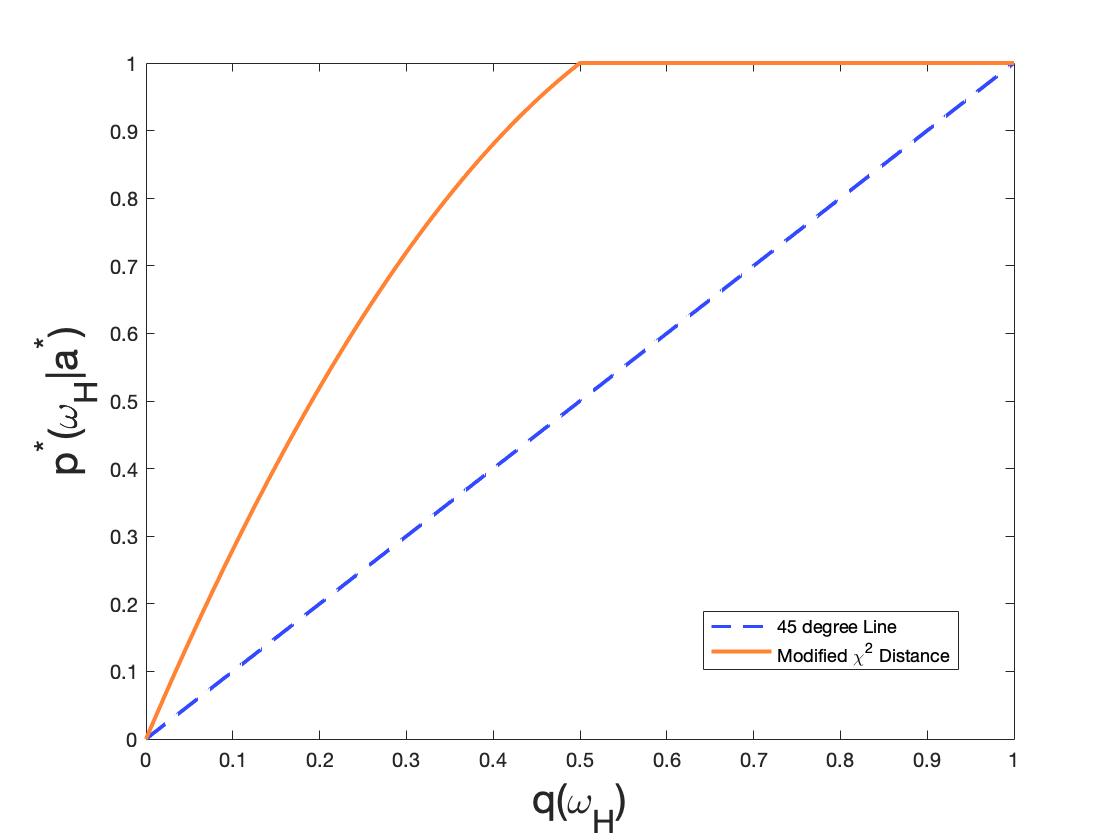}
    \caption{Modified $\chi^2$}
    \label{modchi}
  \end{subfigure}
 
  \begin{subfigure}{6cm}
    \centering\includegraphics[width=5cm]{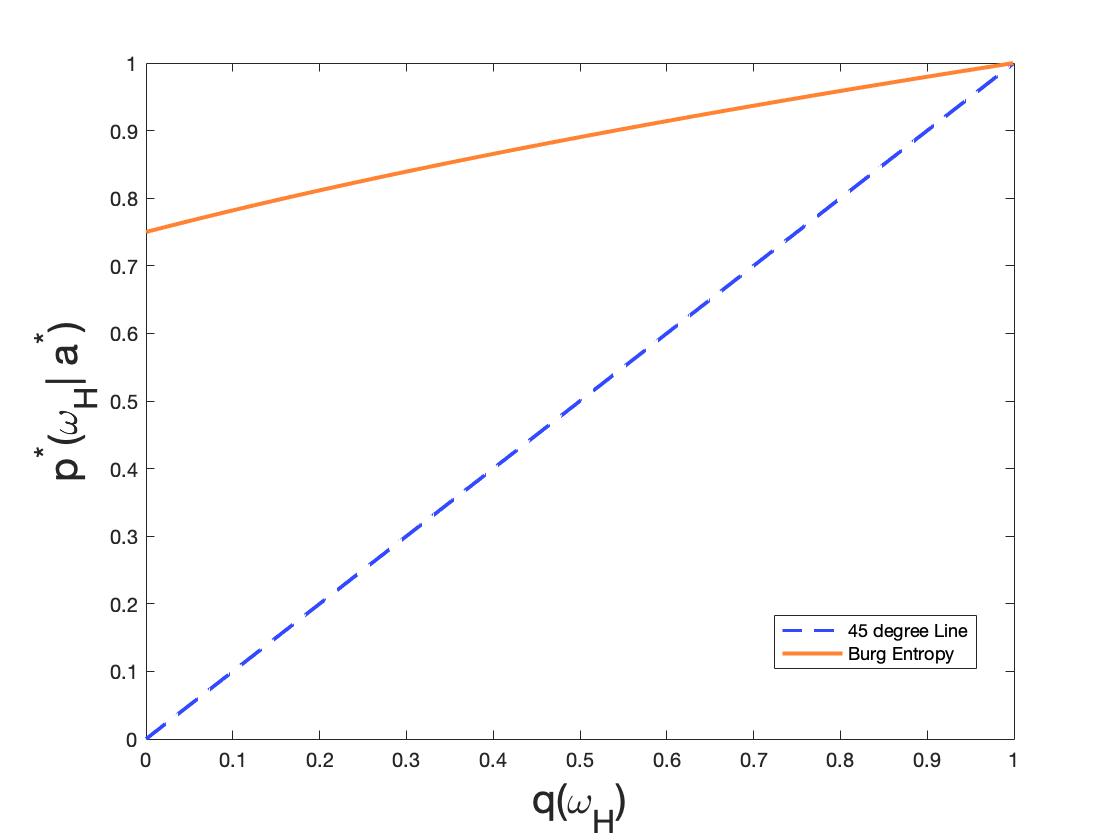}
    \caption{Burg Entropy}
    \label{burg}
  \end{subfigure}
\end{figure}

A mode of comparison, we  first explore  the Kullback-Leibler case. Figure \ref{KLgraph} illustrates the relationship between prior and subjective beliefs in the case of the Kullback-Leibler distance. It is important to note a characteristic of the Kullback-Leibler divergence: $p_{a^\star}^\star(\omega)>0$ if and only if $q(\omega)>0$. This means that the Kullback-Leibler divergence does not generate cognitive censorship.
\smallskip

In the case where the $\phi$-divergence is the modified $\chi^2$ distance, cognitive censorship can be generated by the WT agent.\footnote{The details of this case can be found in \S\ref{Cognitive_censorship_Chi} in Appendix \ref{details_examples}. } Specifically, if $q(\omega_H)\geq 1/2$, the agent censors the state $\omega_L$ and assigns $p_{a^\star}^\star(\omega_H) = 1$ and $p_{a^\star}^\star(\omega_L) = 0$ for the optimal action $a^\star$. This behavior is depicted in Figure \ref{modchi}.

Figure \ref{modchi} illustrates the optimal subjective belief $p^\star(a^\star)$ for different values of prior probability $q(\omega_H)$ in the case of the modified $\chi^2$ distance. It shows that when $q(\omega_H)\geq 1/2$, the WT agent completely disregards the possibility of the low state $\omega_L$ and believes with certainty that the high state $\omega_H$ will occur. This type of belief structure, where a state is completely ignored, and another state is believed with certainty, results from optimism and cannot be generated using the Kullback-Leibler divergence.

This example highlights how different $\phi$-divergences can lead to distinct cognitive behaviors in the WT model. In this case, the modified $\chi^2$ distance allows for cognitive censorship, where the agent selectively ignores certain states based on their prior probabilities.

Our result on cognitive censorship expands in a nontrivial way the analysis  in \cite{mayraz2019priors} and \cite{CaplinLeahy2019} who study the properties of WT behavior in the case of the Kullback-Leibler distance.
\subsection{Cognitive emergence} The optimal belief $p^\star(a^\star)$ exhibits cognitive emergence when for some state $\omega$ , the associated prior probability is $q(\omega) =0$ and $p_{a^\star}^\star(\omega) > 0$. Intuitively, cognitive emergence occurs when a WT agent believes that  an ``impossible'' state $\omega$ is possible.
\smallskip

The following result provides a necessary condition for a DM to exhibit cognitive emergence.
\begin{proposition}\label{popping}
Let $a^\star$ and $p^\star(a^\star)$ be an optimal solution to the WT problem (\ref{wishful_thinking_problem}). Then:
\begin{itemize}
\item[(i)]$p^\star(a^\star)$ can generate cognitive emergence  only if the following condition holds:
\begin{equation}\label{limit_cognitive_emergence}
\lim_{t \rightarrow \infty}\frac{\phi(t)}{t} = b < \infty
\end{equation}
\item[(ii)]If  condition (\ref{limit_cognitive_emergence}) holds, then state $\omega$ can emerge only if $u(a^\star,\omega) = \bar{u}_{a^\star}$, where $\bar{u}_{a^\star}=\max_{\omega^\prime\in \Omega}u(a^\star,\omega^\prime)$.
\item[(iii)]Finally if a state emerges it must hold that $\lambda_{a^\star}^\star = \bar{u}_{a^\star} - b$
\end{itemize}
\end{proposition}

Part (i) of the proposition provides condition (\ref{limit_cognitive_emergence}), which ensures that the cost $C_{\phi}(p^\star(a^\star)\|q)$ remains finite (bounded) when the DM exhibits cognitive emergence. This condition guarantees that the cost function remains well-defined even when $p_{a^\star}^\star(\omega)=0$, allowing for a meaningful analysis of cognitive emergence. Part (ii) establishes that a state $\omega$ can emerge if associated with the highest payoff utility of the optimal choice $a^\star$. This means the DM assigns positive probability to states with the highest utility among all available choices. This behavior captures the DM's preference for skewness, as they are willing to assign positive probability to unlikely states that offer a potentially high utility. This aligns with the findings of \cite{BrunnermeierParker_20005} and \cite{BurghMelo2024b} regarding the preference for skewness in optimistic decision-making.

Part (iii) provides the exact value for $\lambda^\star_{a^\star}$ when cognitive emergence occurs. This value represents the cutoff utility level beyond which the DM assigns positive probability to a state. It characterizes the DM's threshold for cognitive emergence, indicating the point at which the DM starts considering a state as a possibility.

\smallskip

To show how  cognitive emergence operates,  we revisit  the environment where  $\Omega=\{ \omega_H,\omega_L\}$ and $A=\{1,\ldots,n\}$. Let $(a^\star,p^\star(a^\star))$ be an optimal solution where  the state contingent utilities  are $u(a^\star,\omega_H) = 4$ and $u(a^\star,\omega_L) = 0$. Assume that $\phi$ corresponds to the Burg Entropy (see Table \ref{table:1}).  For $q(\omega_H) \in (0,1)$, and applying Lemma \ref{dualprob} and Proposition \ref{Belief_choice_characterization}, we get:
$$p^{\star}_{a^\star}(\omega_H) = \frac{2q(\omega_H)}{\sqrt{9 + 16q(\omega_H)} - 3}$$
Now, we show how the state  $\omega_H$ emerges  when $q(\omega_H)\longrightarrow 0^+$, note that:
$$\lim_{q(\omega_H) \rightarrow 0^+} \frac{2q(\omega_H)}{\sqrt{9 + 16q(\omega_H)} - 3} = \frac{3}{4}$$
Therefore, when $q(\omega_H) = 0$ the DM's optimal belief choice implies  that the state $\omega_H$ emerges with  $p_{a^\star}^\star(\omega_H) = \frac{3}{4}$. Figure \ref{burg} displays the relationship between $q$ and $p^\star(a^\star)$.



From a behavioral standpoint, the previous analysis captures a situation where an optimistic DM ignores the ``impossibility'' of $\omega_H$, and instead, she believes its probability is $p_{a^\star}^\star(\omega_H) = \frac{3}{4}$. 
\smallskip

In simple terms,  the pattern of cognitive emergence establishes that optimism drives the DM to disregard impossibility.  We explore this further in \S\ref{emergence_risky_assets} in the Appendix. In particular, we study a situation where a DM must decide between a highly risky and riskless asset. In this regard, the cognitive emergence can be  useful to explain the recent surge in digital asset scams with promises of exorbitant returns (Cryptocurrencies and NFTs). There are serious doubts about these assets' fundamental market values (\cite{cheah2015speculative}). When the DM is an EU maximizer, experts' opinions (prior beliefs) can dissuade investment in these assets. The Appendix shows this is not necessarily true  when investors display WT behavior. Expert advice can be ignored, and many investors hold highly unrealistic beliefs. In this case, we can say that the DM's behavior is consistent with a preference for skewness. In a different setting,  \cite{BrunnermeierParker_20005} derives a similar result in portfolio allocation problems.

\subsection{Extreme Cognitive Dissonance}\label{actioncont}

In this section, we delve into a fundamental characteristic of motivated reasoning and WT models, drawing a connection between our concept of extreme subjective beliefs and actions. Due to the intimate relationship between actions and beliefs, subjective beliefs are treated as contingent upon the chosen action. Consequently, a DM assigns distinct subjective beliefs to each action throughout the decision-making process.
\smallskip

Action-contingent beliefs have been observed in various contexts. For instance, studies have documented situations where well-educated employees working with hazardous chemicals significantly underestimate the risks associated with their work (\cite{akerlof1982economic}). Similarly, in the context of the COVID-19 pandemic, research has shown that employees' beliefs about the safety of returning to work can be influenced by their motivated reasoning and can align with their preferred course of action (\cite{orhun2021motivated}). These examples illustrate that a DM's beliefs tend to align with their chosen action, and if they were required to take a different action, their beliefs might change accordingly.
\smallskip

Because our model can generate extreme beliefs, it is crucial to emphasize that it enables the association of substantially divergent beliefs with different actions. This becomes especially intriguing when a DM has a personal stake in the outcome (\cite{granberg1988political}). The field of politics provides a fitting illustration, as DMs often maintain significantly disparate beliefs that strongly correspond to their voting patterns. To illustrate this phenomenon within our model, we present the following example:

\begin{example} Let $A=\{a_1,a_2,a_3\}$ and $\Omega=\{\omega_1,\omega_2\}$. The payoff structure is summarized in the following table:

\begin{center}
\begin{tabular}{||c c c c||} 
 \hline
  & $a_1$ & $a_2$ & $a_3$\\ [0.5ex] 
 \hline\hline
 $\omega_1$ & $4$ & $3$ & $0$\\ 
 \hline
 $\omega_2$ & $0$ & $3$ & $4$\\
 \hline
\end{tabular}
\end{center}
\bigskip
Assume that $q(\omega_1)=q(\omega_2)={1\over 2}$ and the cost $C_\phi(p\|q)$ determined by the modified $\chi^2$ distance with $\delta = 1$. Using Lemma \ref{dualprob}, we find $\lambda_{a_1}^* = \lambda_{a_3}^* = 2$ and $\lambda_{a_2}^* = 3$.\footnote{Note that for $a_1$ $(a_3)$ any $\lambda \geq 2$ implies that the DM would censor $\omega_2$ ($\omega_1)$.} The previous fact implies that
$$V_{\phi}(U(a_1)) = V_{\phi}(U(a_2)) = V_{\phi}(U(a_3)) = 3$$
\smallskip

From the preceding equation, we can deduce that the DM is indifferent among $a_1$, $a_2$, and $a_3$. Nevertheless, each action is associated with a distinct subjective probability vector, which can be obtained by applying Proposition \ref{Belief_choice_characterization}. Consequently, there exist three solutions corresponding to the optimal set:

$$ \{(a_1, (1,0)), (a_2, (1/2, 1/2)), (a_3, (0,1))\}.$$\eproof

\end{example}

In the preceding example, the DM censors one state for $a_1$ and $a_3$, whereas for $a_2$, the DM does not distort her beliefs. It is evident in this example that $a_2$ yields the highest EU, yet the DM remains indifferent. This outcome arises because $a_1$ and $a_3$ are associated with more skewed distributions across states.

\section{Related Literature}\label{relatedlit}  Our paper is situated within several strands of literature, addressing various aspects of wishful thinking (WT) in economic decision-making models.
\smallskip

Firstly, our paper is situated within the realm of WT in economic decision-making models. Several related studies provide an axiomatic foundation for WT behavior, such as those by \cite{mayraz2019priors} and \cite{kovach2020twisting}. Additionally, \cite{bracha2012affective} and \cite{CaplinLeahy2019} share similarities with our research and are closely related to our study. Like us, they consider a decision-making model where the DM selects a probability distribution over states based on the associated EU and the cost of distorting baseline beliefs. \cite{CaplinLeahy2019} quantify this cost using only the Kullback-Leibler distance between subjective and prior beliefs. \cite{bracha2012affective} consider a general cost function while modeling decision-making as a game between a ``subjective'' process and ``objective'' process. However, several crucial distinctions exist between \cite{bracha2012affective} and \cite{CaplinLeahy2019} and our results. In particular, with respect to \cite{CaplinLeahy2019}, our framework encompasses the Kullback-Leibler distance as a specific case within a broader framework of WT. This allows us to explore cognitive emergence and censorship, the connection between WT and risk measures. These aspects go beyond the scope of \cite{CaplinLeahy2019}. Regarding \cite{bracha2012affective}, our analysis focuses on the class of statistical distances given by the $\phi$-divergence functions, whereas they do not focus on specific cost functions. This distinction results in \cite{bracha2012affective} being unable to characterize WT beliefs in terms of risk measures, generate censoring and emergence, or characterize optimal beliefs in closed form, as our framework allows.

Thus, while our work shares common ground with these two papers, our approach extends beyond their respective frameworks, offering insights into the complexities of WT behavior and its implications for decision-making processes.

\smallskip

Secondly, our paper is closely related to the literature on optimal expectations, particularly the work of \cite{BrunnermeierParker_20005}. Their study delves into a dynamic model of belief choice, although it does not explicitly consider the cost associated with maintaining optimistic beliefs.

While there are similarities between their model and ours, a fundamental structural difference lies in how we explicitly model the cost of distorted beliefs. This explicit consideration enables us to characterize optimal beliefs and uncover phenomena such as cognitive censorship, cognitive emergence, and the connection between Wishful Thinking (WT) behavior and risk measures. These aspects are not explicitly addressed in the model presented by \cite{BrunnermeierParker_20005}.

By explicitly integrating the cost associated with biased beliefs into our model, we provide insights into the complexities of decision-making under optimism, offering a nuanced understanding of how individuals navigate optimistic belief formation and its consequences.

\smallskip

Finally, our paper is also related to the literature on robustness in economic models.  Specifically,  \cite{Hansen_Sargent_2001, Hansen_Sargent2008} introduce a robustness approach to address the concern of model misspecification. They adopt a max-min approach, akin to multiple priors models as in \cite{GILBOA1989141} and \cite{maccheroni2006ambiguity}, to make decisions under ambiguity. While robustness and ambiguity models typically take a pessimistic approach to decision-making, our paper focuses on WT behavior and adopts an optimistic approach by studying a max-max problem. By doing so, our paper draws on the active and rapidly growing literature on distributionally robust optimization problems \cite{Shapiro} and \cite{Kuhn_et_al_2019}. This framework provides a powerful tool to handle decision problems under uncertainty, allowing for a consideration of a range of possible distributional assumptions. In our case, we leverage this framework to capture WT behavior and its implications for decision-making.
\smallskip

Thus, while our paper is related to the robustness literature, it differs in terms of its focus on WT, adopting an optimistic approach, and using distributionally robust optimization techniques. These distinctions enable us to study the specific biases and preferences associated with WT and its impact on decision outcomes.
 
\section{Conclusions}\label{concusions}
In this paper, we present a tractable model of WT and optimism, which incorporates the costs and benefits associated with biased beliefs. By doing so, we establish connections between WT behavior and risk measures, demonstrating how optimistic agents tend to opt for riskier alternatives. Our model effectively captures extreme belief behaviors such as cognitive censorship and emergence.
\smallskip

Moving forward, there exist several avenues for extending and enriching our model. One crucial direction involves exploring dynamic environments where decision-making unfolds over multiple stages. Analyzing how WT behavior evolves over time can yield valuable insights into biased belief formation and decision-making dynamics.
\smallskip

Furthermore, our current model primarily focuses on single-agent WT behavior. Extending the analysis to include strategic interactions and WT could offer a deeper understanding of how optimism and biased beliefs shape strategic decision-making and outcomes in interactive settings. For instance, our approach and findings could prove insightful in studying Bayesian persuasion when the receiver exhibits wishful thinking. Previous work on the Kullback-Leibler cost function in this context has been explored by \cite{Augias2023persuading}.
\smallskip

Lastly, experimental testing of the theoretical predictions derived from our model can serve to validate and refine the insights we have obtained. Such experimentation would further contribute to the robustness and applicability of our theoretical framework.

\appendix
\section{Proofs}\label{AppendixA}

\subsection*{Proof of Lemma \ref{dualprob}}

In the optimization problem
$$
V_\phi(U(a))=\max_{p\in\Delta(\Omega)}\left\{\sum_{\omega\in \Omega} p(\omega) u(a,\omega)-\delta D_\phi(p\| q)\right\},
$$
the objective is concave in $p$, and the constraints are linear. Therefore, the optimal value is equal to the optimal value of the Lagrangian dual problem,
\begin{eqnarray}
\mathcal{L}(\lambda)	&=& 
	\inf_{\lambda_a} \max_{p\in \Delta(\Omega)}\left\{\sum_{\omega\in \Omega} p(\omega) u(a,\omega)-\delta D_\phi(p\|q)+\lambda_a\left(1-\sum_{\omega\in \Omega} p(\omega)\right)\right\}\nonumber\\
	&=&\inf _{\lambda_a}\left\{\lambda_a+\sum_{\omega\in \Omega} \sup_{p(\omega) \geq 0}\left\{p(\omega)\left(u(a,\omega)-\lambda_a\right)-\delta\sum_{\omega\in \Omega} q(\omega) \phi\left(\frac{p(\omega)}{q(\omega)}\right)\right\}\right\} \nonumber\\
	&=& \inf_{\lambda_a}\left\{\lambda_a+\sum_{\omega\in \Omega} \sup_{p(\omega) \geqslant 0}\left\{p(\omega)\left(u(a,\omega)-\lambda_a\right)-\delta\sum_{\omega\in \Omega} q(\omega) \phi\left(\frac{p(\omega)}{q(\omega)}\right)\right\}\right\}\nonumber\\
	&=&\inf_{\lambda_a}\left\{\lambda_a+\sum_{\omega\in \Omega} q(\omega) \max_{p(\omega)\geqslant 0}\left\{\frac{p(\omega)}{q(\omega)}\left(u(a,\omega)-\lambda_a\right)-\delta\phi\left(\frac{p(\omega)}{q(\omega)}\right)\right\}\right\}\nonumber\\
	&=&\inf_{\lambda_a}\left\{\lambda_a+\delta\sum_{\omega\in \Omega} q(\omega) \sup _{t \geqslant 0}\left\{t\left((u(a,\omega)-\lambda_a)/\delta\right)-\phi(t)\right\}\right\}\nonumber\\
	&=&\inf_{\lambda_a}\{ \lambda_a+\delta\mathbb{E}_{q}(\phi^*((u(a,\omega)-\lambda_a)/\delta))\}\label{inf_is_min}
\end{eqnarray}

Now, solving for $\lambda_a$ we find that there exists a unique $\lambda_a^\star$ that satisfies 
\begin{equation}\label{lambdastar}
    \sum_{\omega\in \Omega}\phi^{\ast\prime}((u(a,\omega)-\lambda_a^\star)/\delta)q(\omega)=1.
\end{equation}

Furthermore, $\lambda_a^\star\in [\underline{u}_a,\bar{u}_a]$. To see this, recall that $\phi^{*\prime}$ is monotonically increasing and $\phi^{*\prime}(0) = 1$. Assume that there exists an optimal $\tilde{\lambda}_a$ such that $\tilde{\lambda}_a>\bar{u}_a$, then $\sum_{\omega\in \Omega}\phi^{\ast\prime}((u(a,\omega)-\tilde{\lambda}_a\
)/\delta)q(\omega)<1$. Therefore $\tilde{\lambda}_a$ cannot be optimal. Now assume there exists an optimal $\tilde{\lambda}_a$ such that $\tilde{\lambda}_a<\underline{u}_a$, then $\sum_{\omega\in \Omega}\phi^{\ast\prime}((u(a,\omega)-\tilde{\lambda}_a)/\delta)q(\omega)>1$. Therefore $\tilde{\lambda}_a$ cannot be optimal. We conclude that it must hold $\lambda_a^\star\in [\underline{u}_a,\bar{u}_a]$. Then the infimum  in (\ref{inf_is_min}) is achieved, and we can write
$$\max_{p\in \Delta(\Omega)}\left\{\sum_{\omega\in \Omega} p(\omega)u(a,\omega)-\delta D_\phi(p\| q)\right\}=\min_{\lambda_a\in[\underline{u}_a,\bar{u}_a]}\{ \lambda_a+\delta\mathbb{E}_{q}(\phi^*((u(a,\omega)-\lambda_a)/\delta))\}.$$

The uniqueness of $\lambda_a^\star$ follows from Assumption \ref{phi_diffferentiable_Assumption}, which implies that problem (\ref{equivalence_belief_choice}) is strictly convex in $\lambda_a$. \eproof

\subsection*{Proof of Proposition \ref{Belief_choice_characterization}}

From Lemma \ref{dualprob}, we know that  the necessary  and sufficient first-order conditions in problem (\ref{equivalence_belief_choice}) yields $p_a^\star(\omega)=\phi^{\ast\prime}((u(a,\omega)-\lambda_a^\star)/\delta)q(\omega)$  for all $a\in A$, $\omega\in \Omega$. Then by a straightforward application of the  envelope theorem, we get ${\partial V_\phi(U) \over\partial u(a,\omega) }=\phi^{*\prime}((u(a,\omega)-\lambda_a^\star)/\delta )q(\omega)$ for all $\omega\in \Omega$. Thus we conclude $\nabla V_\phi(U(a))=p^\star(a)$ for all $a\in A$.\eproof

\subsection*{Proof of Corollary \ref{higherstate}}

From expression (\ref{Optimal_beliefs}) we know that the  weights   $\phi^{\ast\prime}((u(a,\omega)-\lambda_a^\star)/\delta)$ are increasing in  $u(a,\omega)$.  To see the bias, we note that  when $u(a,\omega)>u(a,\omega^\prime)$, the strict monotonicity of the gradient 
$\phi^{\ast\prime}$ implies that $\phi^{\ast\prime}((u(a,\omega)-\lambda_a^\star)/\delta)>\phi^{\ast\prime}((u(a,\omega^\prime)-\lambda_a^\star)/\delta)$, which implies
$p_a^\star(\omega)>p_a^\star(\omega^\prime)$.

Next, we note $p_a^\star(\omega)=\phi^{\ast\prime}((u(a,\omega)-\lambda_a^\star)/\delta)q(\omega)$ is linearly increasing in $q(\omega)$. Combining these facts implies that given states $\omega$ $\omega^\prime$ with $q(\omega) > q(\omega^\prime)$  and $u(a,\omega)>u(a,\omega^\prime)$, we have $p_a^\star(\omega)>p_a^\star(\omega^\prime)$.

\eproof

\subsection*{Proof of Proposition \ref{Vproperties}}

Proofs of (i)-(iv)
  
(i) For any $\phi\in \Phi$ and any $c \in \mathbb{R}$
$$
\begin{aligned}
V_\phi(U(a)+c) & =\inf _{\lambda_a \in \mathbb{R}}\{\lambda_a+\mathbb{E} (\phi^\ast(u(a,\omega)+c-\lambda_a))\} \\
& =c+\inf _{\lambda_a \in \mathbb{R}}\{\lambda_a-c+\mathbb{E} (\phi^\ast(u(a,\omega)-(\lambda_a-c)))\}=c+V_\phi(U(a)) .
\end{aligned}
$$

(ii) Since $\phi\in \Phi$, then $\phi^\ast(0)=0,0 \in \partial \phi^\ast(1)$ and the convexity  of $\phi^\ast$ implies $\phi^\ast(t) \geq t$, and hence
$$
V_\phi(U(a))\geq \inf _{\lambda_a \in \mathbb{R}}\{\lambda_a+(c-\lambda_a)\}=c .
$$
For the converse inequality, since $\phi^\ast(0)=0$, one has $V_\phi(c) \leq\{c+\phi^\ast(c-c)\}=c$. Then we conclude that $V_\phi(c)=c$.\\

(iii) If $U(a) \leq\tilde{U}(a)$, then $U(a)-\lambda_a \leq \tilde{U}(a)-\lambda_a$, and since $\phi^\ast$ is non-decreasing it follows that,
$$
V_\phi(U(a))=\inf _{\lambda_a\in \mathbb{R}}\{\lambda_a+\mathbb{E} \phi^\ast(u(a,\omega)-\lambda_a))\} \leq \inf _{\lambda_a \in \mathbb{R}}\{\lambda_a+\mathbb{E} \phi^\ast(\tilde{u}(a,\omega)-\lambda_a))\}=V_\phi(\tilde{U}(a)) .
$$

(iv) Let $\alpha \in(0,1)$ and for any random variables $U_1(a), U_2(a)$, let $U_\alpha(a):=\alpha U_1(a)+(1-\alpha) U_2(a)$. Since $\phi^\ast$ is convex, the function $f(z, \lambda_a):=\lambda_a+\phi^\ast(z-\lambda_a)$ is jointly convex over $\mathbb{R} \times \mathbb{R}$. Therefore, for any $\lambda_a^1, \lambda_a^2 \in \mathbb{R}$, and with $\lambda_a^\alpha\triangleq \alpha\lambda_a^1+(1-\alpha)\lambda_a^2$, one has,
$$
\mathbb{E} f\left(U_\alpha(a), \lambda_\alpha\right) \leq \lambda \mathbb{E} f\left(U_1(a), \lambda_a^1\right)+(1-\alpha) \mathbb{E} f\left(U_2(a), \lambda_a^2\right) .
$$
Since $V_\phi\left(U_\alpha(a)\right)=\inf _{\lambda_a \in \mathbb{R}} \mathbb{E} f\left(U_\alpha(a), \lambda\right)$, it follows that,
$$
\begin{aligned}
V_\phi\left(U_\alpha(a)\right) & \leq \inf _{\lambda_a^1, \lambda_a^2}\left\{\alpha \mathbb{E} f\left(U_1(a), \lambda_a^1\right)+(1-\alpha) \mathbb{E} f\left(U_2(a), \lambda_a^2\right)\right\} \\
& =\alpha V_\phi\left(U_1(a)\right)+(1-\alpha) V_\phi\left(U_2(a)\right) .
\end{aligned}
$$\eproof

\subsection*{Proof of Proposition \ref{euequiv}}
The WT agent's optimal actions are given by:
\begin{eqnarray}
A_{eu}(\tilde{U})\nonumber	&=& 
	\arg\max_{a \in A}\mathbb{E}_{q}(\tilde{u}(a,\omega)),\nonumber\\
	&=&\arg\max_{a \in A}[\lambda_a^\star+\delta\mathbb{E}_{q}(\phi^*((u(a,\omega)-\lambda_a^\star )/\delta))] \text{(By Definition)},\nonumber\\
	&=&\arg\max_{a \in A}\max_{p\in\Delta(\Omega)}\left[\sum_{\omega\in \Omega}p(\omega)u(a,\omega)-\delta C_{\phi}(p\| q)\right],\nonumber\\
    &=&A_{wt}(U),\nonumber
\end{eqnarray}
where the last equality follows from the definition of $A_{wt}(U)$.

\eproof

\subsection*{Proof of Corollary \ref{saddleWT}}

Thanks to Lemma \ref{wishful_thinking_problem}, we know that for each  $a\in A$ 
$$V_{\phi}(U(a)) = \max_{p \in \Delta(\Omega)} \sum_{\omega \in \Omega} p(\omega) u(a,\omega) - \delta D_\phi(p\|q)=\min_{\lambda_a\in \Lambda(a)}\Psi(a,\lambda_a).$$
 Then it follows that  problem (\ref{wishful_thinking_problem}) is equivalent to
$$\max_{a\in A}\min_{\lambda_a\in \Lambda(a)}\Psi(a,\lambda_a).$$\eproof

\subsection*{Proof of Proposition \ref{zerochoice}}

To study a situation of cognitive censorship, we consider a state $\hat{\omega}$ such that $q(\hat{\omega}) > 0$. Let $a^\star$ and $p^\star(a^\star)$ be an optimal solution to WT problem (\ref{wishful_thinking_problem}).

From Proposition \ref{Belief_choice_characterization}, the subjective belief vector is given by $p_{a^\star}^\star(\hat{\omega})=\phi^{\ast\prime}(u(a^\star,\hat{\omega})-\lambda_{a^\star}^\star)q(\hat{\omega})$ where $s_{\hat{\omega}}^\star = u(a^\star,\hat{\omega})-\lambda_{a^\star}^\star$. Because $q(\hat{\omega}) > 0$, $ p_{a^\star}^\star(\hat{\omega}) = 0$ if and only if $\phi^{\ast\prime}(u(a^\star,\hat{\omega})-\lambda_{a^\star}^\star) = 0$. For $\phi^{\ast\prime}(u(a^\star,\hat{\omega})-\lambda_{a^\star}^\star) = 0$, it must be the case that $\phi^{\ast}(u(a^\star,\hat{\omega})-\lambda_{a^\star}^\star) = c$. $\phi^*$ is a monotone, non-decreasing function, so there exists $z$ such that  $\phi^{\ast}(u(a^\star,\hat{\omega})-\lambda_{a^\star}^\star + z) = c$ only if $\lim_{z\rightarrow -\infty} \phi^\ast(u(a^\star,\hat{\omega})-\lambda_{a^\star}^\star + z) = c > -\infty$.

Therefore it is possible to have $p_{a^\star}^\star(\hat{\omega}) = 0$ only if $$\lim_{z \rightarrow -\infty} \phi^\ast(u(a^\star,\hat{\omega})-\lambda_{a^\star}^\star+ z) = c > -\infty$$.

We first prove condition (i). $\lim_{t \rightarrow 0+}\phi(t) < \infty$, is a necessary condition for cognitive censorship by contradiction. Assume $\lim_{t \rightarrow 0+}\phi(t) = \infty$. Applying the definition of a convex conjugate, this implies $\lim_{z \rightarrow -\infty} \phi^*(u(a^\star,\hat{\omega})-\lambda_{a^\star}^\star + z) = -\infty$. Therefore, $p_{a^\star}^\star(\hat{\omega}) = 0$ cannot hold if $\lim_{t \rightarrow 0+}\phi(t) = \infty$. This implies $\lim_{t \rightarrow 0+}\phi(t) < \infty$ is a necessary condition for cognitive censorship. This proves Condition (i).
\smallskip

We now prove condition (ii). We achieve this by contradiction. Assume $\lim_{t \rightarrow 0+}\phi^\prime(t) = -\infty$. \cite{LoveBayraksan_2015} shows that this implies $ \lim_{z \rightarrow - \infty}\phi^\ast(u(a^\star,\hat{\omega})-\lambda_{a^\star}^\star + z)= c$ asymptotically, but there does not exist $z$ such that $\phi^*(u(a^\star,\hat{\omega})-\lambda_{a^\star}^\star + z)= c$. Therefore for all $\hat{\omega}$ it must hold $\phi^{*\prime}(u(a^\star,\hat{\omega})-\lambda_{a^\star}^\star) > 0$. This implies $p_{a^\star}^\star(\hat{\omega}) > 0$. Therefore, $p_{a^\star}^\star(\hat{\omega}) = 0$ cannot hold if $\lim_{t \rightarrow 0+}\phi'(t) = -\infty$. This  implies $\lim_{t \rightarrow 0+}\phi'(t) > -\infty$ is a necessary condition for cognitive censorship. We have proved condition (ii).
\smallskip

To prove the final assertion, note that $\phi^*$ is a monotone, non-decreasing, continuous, convex function. Then, if there exists a $ z$ s.t. $\phi^{\ast\prime}(u(a^\star,\hat{\omega})-\lambda_{a^\star}^\star + z) = 0$, there exists a cutoff $\hat{z}_{a^*}$ such that $\phi^{\ast\prime}(u(a^\star,\hat{\omega})-\lambda_{a^\star}^\star + z) = 0$ if $z < \hat{z}_{a^*}$ and $\phi^{\ast\prime}(u(a^\star,\hat{\omega})-\lambda_{a^\star}^\star + z) < 0$ if $z > \hat{z}_{a^*}$. Equivalently, there exists $\tilde{u}(a^\star)$ such that $\phi^{\ast\prime}(u(a^\star,\hat{\omega}) - \lambda^\star_{a^\star}) = 0$ if and only if $u(a^\star,\hat{\omega}) \leq \tilde{u}(a^\star)$. Finally, we can see $p_{a^\star}^\star(\hat{\omega}) = 0$ if and only if $u(a^\star,\hat{\omega}) \leq \tilde{u}(a^\star)$. This proves the final assertion.

\eproof

\subsection*{Proof of Proposition \ref{popping}}

To study a situation of cognitive emergence, assume there exists a state $\hat{\omega}$ such that $q(\hat{\omega}) = 0$. Let $a^\star$ and $p^\star(a^\star)$ be an optimal solution to WT problem (\ref{wishful_thinking_problem}).

Recall that we defined $0\phi(\frac{c}{0}) = c lim_{t \rightarrow \infty} \frac{\phi(t)}{t}$ $\forall c> 0$ and $0\phi(\frac{0}{0}) = 0$.

We first prove (i), $\lim_{t \rightarrow \infty}\frac{\phi(t)}{t} = b < \infty$ is a necessary condition for cognitive emergence by contradiction. Assume $\lim_{t \rightarrow \infty}\frac{\phi(t)}{t} = \infty$. Using the definition of $V_\phi(U(a^\star))$ we know that:
$$\inf_{\lambda_{a^\star}}\left\{\lambda_{a^\star}+ \sum_\omega \sup _{p_{a^\star}(\omega) \geqslant 0}\left\{p_{a^\star}(\omega)\left(u(a^\star,\omega)-\lambda_{a^\star}\right)-q(\omega) \phi\left(\frac{p_{a^\star}(\omega)}{q(\omega)}\right)\right\}\right\}$$

Consider the term inside the summation corresponding to state $\hat{\omega}$:

$$\sup _{p_{a^\star}(\hat{\omega}) \geqslant 0}\left\{p_{a^\star}(\hat{\omega})\left(u(a^\star,\hat{\omega})-\lambda_a^\star\right)-
0 \phi\left(\frac{p_{a^\star}(\hat{\omega})}{0}\right)\right\} = \sup_{p_{a^\star}(\hat{\omega})\geq 0}\{F(p_{a^\star}(\hat{\omega}))\}$$

where:$$F(p_{a^\star}(\hat{\omega})) = 
\begin{cases}
-\infty & \text{ if } p_{a^\star}(\hat{\omega}) >0\\
0 & \text{ if } p_{a^\star}(\hat{\omega}) = 0
\end{cases}$$

This yields $p_{a^\star}^\star(\hat{\omega}) = 0$. Then $p_{a^\star}^\star(\hat{\omega}) > 0$ cannot hold if $\lim_{t \rightarrow \infty}\frac{\phi(t)}{t} = \infty$. We can then say $\lim_{t \rightarrow \infty}\frac{\phi(t)}{t} = b < \infty$ is a necessary condition for cognitive emergence.

To prove assertions (ii) and (iii), let $\lim_{t \rightarrow \infty}\frac{\phi(t)}{t} = b < \infty$. Again consider the term inside the summation associated with state $\hat{\omega}$:

$$\sup _{p_{a^\star}(\hat{\omega}) \geqslant 0}\left\{p_{a^\star}(\hat{\omega})\left(u(a^\star,\hat{\omega})-\lambda^\star_{a^\star}\right)-
0 \phi\left(\frac{p_{a^\star}(\hat{\omega})}{0}\right)\right\}$$

$$\sup _{p_{a^\star}(\hat{\omega}) \geqslant 0}\left\{p_{a^\star}(\hat{\omega})\left(u(a^\star,\hat{\omega})-\lambda^\star_{a^\star} - b\right)\right\}$$

The optimal choice depends on the sign of $(u(a^\star,\hat{\omega})-\lambda^\star_{a^\star} - b)$. Consider all three cases:

\begin{enumerate}
    \item $(u(a^\star,\hat{\omega})-\lambda^\star_{a^\star} - b) > 0$: The supremum yields $p_{a^\star}^\star(\hat{\omega}) = \infty$, making the term unbounded. Because $\lambda^\star_{a^\star}$ is an argument which minimizes it's objective function, $\lambda^\star_{a^\star}$ cannot admit this case
    \item $(u(a^\star,\hat{\omega})-\lambda^\star_{a^\star} - b) < 0$. The supremum yields $p_{a^\star}^\star(\hat{\omega}) = 0$. 
    \item $(u(a^\star,\hat{\omega})-\lambda^{\star}_{a^\star} - b) = 0$. There is no restriction on $p_{a^\star}^\star(\hat{\omega})$, i.e. $p_{a^\star}^\star(\hat{\omega})\in [0,\infty)$
    
\end{enumerate}

Therefore, only cases (2) and (3) will be possible with the optimal $\lambda_{a^\star}^\star$: $(u(a^\star,\omega)-\lambda_{a^\star}^\star - b) \leq 0$ for all states $\omega \in \Omega$. The state $\hat{\omega}$ can have subjective probability $p_{a^*}^\star(\hat{\omega})$ only if $u(a^\star,\hat{\omega})-\lambda_{a^\star}^\star - b = 0$. In other words: $\lambda_{a^\star}^\star = u(a^\star,\hat{\omega}) - b$. Furthermore, it must hold that $u(a^\star,\hat{\omega}) = \bar{u}(a^\star)$; otherwise, case (1) would arise for another state. This proves assertions (ii) and (iii).\eproof

\section{Details of cognitive censorship and emergence}\label{details_examples}

\subsection{Modified $\chi^2$ and cognitive censorship}\label{Cognitive_censorship_Chi} We now provide the details  of  the modified $\chi^2$ distance  generating $p_{a^\star}^\star(\omega_H)=1.$ We recall  that $\Omega=\{ \omega_H,\omega_L\}$ and $A=\{1,\ldots,n\}$. Let $(a^\star,p^\star(a^\star))$ be an optimal solution where  the state contingent utilities  are $u(a^\star,\omega_H) = 4$ and $u(a^\star,\omega_L) = 0$.  To determine the optimal subjective belief vector $(p_{a^\star}^\star(\omega_H), p_{a^\star}^\star(\omega_L))$, we apply Lemma \ref{dualprob} to find:
$$\lambda_{a^\star}^\star = \begin{cases}
    6 - \frac{2}{q_H} & q_H \geq 1/2\\
    4q_H & q_H < 1/2
\end{cases}$$
    Notice that $\lambda_{a^\star}^\star \geq 2 \rightarrow p_{a^\star}^\star(\omega_L) = 0$ as $\phi^{*\prime}(\lambda_{a^\star}^\star) = 0$. 
     Then the optimistic DM sets   $p_{a^\star}^\star(\omega_H) = 1$ and $p_{a^\star}^\star(\omega_L) = 0$ if $q(\omega_H) \geq \frac{1}{2}$. For  $q(\omega_H) < \frac{1}{2}$, apply Proposition \ref{Belief_choice_characterization} to recover subjective probabilities.

\subsection{Cognitive emergence and risky assets}\label{emergence_risky_assets}

Consider a DM with access to a highly risky asset $a_R$ that returns a high payoff  of 4 in  the high state $\omega_H$ with probability $q(\omega_H) = 0$ (or arbitrarily close). Otherwise, the asset returns 0 in low state $\omega_L$ ($q(\omega_L)$ = 1). The DM also has access to a safe asset $a_S$, which returns a guaranteed normalized value of 1. The prior probability tells us there is little chance the risky asset will be a good investment. Assume that the  DM  has a cost $C_\phi(p\|q)$ determined by the Burg divergence with $\delta = 1$. From our analysis of emergence, we know that the DM's optimal beliefs imply that $p_{a_R}^\star(\omega_H) = \frac{3}{4}$. Using this fact, it follows that
$$V(U(a_R)) = 3- log4 > V(U(a_S)) = 1$$
Accordingly, the optimal solution is:
$$(a^\star, p^\star(a^\star)) = (a_R, (3/4,1/4))$$
In this example, cognitive emergence implies that the DM will purchase the risky asset.

\bibliographystyle{plainnat} 
\bibliography{bibliography.bib}

\end{document}